\documentclass[twocolumn,prl,superscriptaddress,aps]{revtex4-2}
\usepackage{amsmath}
\usepackage{graphicx}
\usepackage{siunitx}
\usepackage{ulem}
\DeclareSIUnit\gauss{G}
\DeclareSIUnit\bohr{a_{B}}

\usepackage{hyperref}

\graphicspath{{../}}

\usepackage{color}
\definecolor{mygreen}{rgb}{0,0.5,0} 
\definecolor{mygrey}{rgb}{0.5,0.5,0.5} 
\definecolor{myred}{rgb}{0.75,0,0} 
\definecolor{myblue}{rgb}{0,0,0.75} 
\definecolor{mymagenta}{cmyk}{0,1,0,0.12} 
\definecolor{mycyan}{cmyk}{1,0,0,0.12} 
\definecolor{myorange}{rgb}{1,0.5,0}  
\definecolor{myviolet}{rgb}{0.5,0.3,1} 
\definecolor{mybrown}{rgb}{0.542969,0.269531, 0.0742188} 

\usepackage{ulem}

\newcommand{\btext}[1]{{\color{myblue}#1}}
\renewcommand{\btext}[1]{#1}

\newcommand{\supmax}{^{(\mathrm{max})}}

\begin{document}
\newcommand{\microtrap}{microtrap~}
\newcommand{\microtraps}{microtraps~}
\newcommand{\gammaNat}{\gamma_0}
\newcommand{\gammaJump}{\Gamma}
\newcommand{\nuprobe}{\nu_\mathrm{pr}}
\newcommand{\Iprobe}{I_\mathrm{pr}}
\newcommand{\MAtom}{M_A}
\newcommand{\TAtom}{T_A}

\newcommand{\mytitle}{\btext{Quantum jump spectroscopy of a single neutral atom for precise sub-wavelength intensity measurements}}
\title{\mytitle}

\newcommand{\ICFO}{ICFO - Institut de Ciencies Fotoniques, The Barcelona Institute of Science and Technology, 08860 Castelldefels, Barcelona, Spain}
\newcommand{\ICREA}{ICREA - Instituci\'{o} Catalana de Recerca i Estudis Avan{\c{c}}ats, 08010 Barcelona, Spain}
\newcommand{\ZH}{State Key Laboratory of Modern Optical Instrumentation, College of Optical Science and Engineering, Zhejiang University, Hangzhou 310027,China}

\author{Lorena C. Bianchet}
\affiliation{\ICFO}
\author{Natalia {A}lves}
\affiliation{\ICFO}
\author{Laura Zarraoa}
\affiliation{\ICFO}
\author{Tomas Lamich}
\affiliation{\ICFO}
\author{Vindhiya Prakash}
\affiliation{\ICFO}
\author{Morgan W. Mitchell}
\affiliation{\ICFO}
\affiliation{\ICREA}

\begin{abstract}
\btext{We present precise, sub-wavelength optical intensity measurement using a single trapped \textsuperscript{87}Rb  atom as a sensor. The intensity is measured by the scalar ac Stark shift it produces on 
the  $F=1 \rightarrow F'=2$ hyperfine transition of the D\textsubscript{2} line,  chosen for its $F' = F+1$ structure and very small tensor polarizability. To boost signal and reduce measurement-induced perturbations, we use a quantum jump spectroscopy technique in which a single absorbed photon on a transition of interest induces the scattering of hundreds of photons on a bright closed transition. The method greatly reduces systematic effects associated with the atomic state, optical polarization, probe power, and atom heating, and  gives the atomic temperature as a second spectroscopic observable. We demonstrate the method by measuring the intensity at the focus of an optical tweezer.  }
\end{abstract}

\maketitle

Individual trapped ions and neutral atoms  can be positioned with sub-micrometer precision, and have been used to detect a variety of environmental perturbations, including static \cite{RusterPRX2017} and oscillating \cite{KotlerN2011} magnetic fields, static electric fields \cite{BiercukNN2010}, and microwaves \cite{WahnschaffeAPL2017}. 
Measurement of optical intensity, which in many scenarios varies on micrometer scales, is a natural application for such sensors \cite{GuthohrleinN2001, DeistPRL2022}.
Sub-wavelength (a.k.a. super-resolving) measurement of both resonant  \cite{GuthohrleinN2001} and off-resonant light \cite{ShihPRA2013, DeistPRL2022} have been demonstrated. Single atoms and ions are also ideal for metrology referenced to unchanging atomic properties, e.g., polarizabilites that can be calculated with high precision \cite{SafronovaPRA2004, SimonCoopLightshift2017}. Single trapped atoms thus offer a route to precision radiometry with high spatial resolution.

Off-resonance light, which for any given atom constitutes the vast majority of the optical spectrum, can be detected by the ac Stark shifts it produces on observable spectral lines. For example, single neutral \textsuperscript{87}Rb atoms in far-off-resonance traps (FORTs) have been used to quantify ac Stark shifts by monitoring fluorescence on the   $F=2\rightarrow F'=3$ cooling transition of the D\textsubscript{2} line  \cite{ShihPRA2013, KaufmanS2014, DeistPRL2022}. While this strong, closed transition is convenient, it is not naturally suited for precision \btext{intensity} measurement {because it, like other strong closed transitions,} has large vector and tensor polarizabilities. This implies 1) that a scalar ac Stark shift to be detected will necessarily be accompanied by a broadening or splitting of the resonance fluorescence line \cite{ShihPRA2013} and 2) that the resonance fluorescence intensity will depend on the polarization of the excitation light, the atomic Zeeman state, and, via the FORT intensity distribution, also the atomic position. These atomic attributes are  easily perturbed by the resonance fluorescence process itself, which modifies the atom’s internal state through optical pumping, and its position through recoil effects \cite{ShihPRA2013}. All these factors complicate the interpretation of the acquired spectra.

Here we introduce a single-atom probing method that greatly reduces these systematic effects, through the use of an open transition and ``quantum jump'' readout \cite{SchmidtS2005, MyersonPRL2008, BascheN1995, GleyzesN2007, VamivakasN2010} to amplify the resulting signal at very low probe power levels.
We apply the  technique to measure the intensity distribution seen by an atom in an optical tweezer, i.e., a strongly-focused FORT \cite{SchlosserN2001}, of the sort used to study quantum light-matter interactions \cite{ChinNC2017, AljunidPRL2009, LeongNC2016, AsenjoGarciaPRX2017, PerczelPRL2017, AlbrechtARX2018}, non-classical atom interference effects \cite{KaufmanS2014, LesterPRL2018, KaufmanN2015,GlicensteinPRL2020}, Rydberg-atom-based quantum information processing \cite{SaffmanRMP2010},  quantum simulation \cite{BernienN2017, LabuhnN2016} \btext{and computation \cite{SparPRL2022, DeistPRL2022, WilsonPRL2022}, manipulation of cold molecules for quantum information and searches for physics beyond the standard model
\cite{Burchesky2021,Cairncross2021,Caldwell2020,Weyland2021}, and also
optomechanics and quantum optomechanics with levitated nanoparticles \cite{Delic2019,Schaefer2021,Windey2019}}. In this application, the method reveals both the trap-center intensity with high precision, and also the atom temperature, both of which are subject to considerable systematic uncertainty when measured by other methods \cite{BianchetORE2021}.

\begin{figure}[t]
\begin{center}
\includegraphics[width=0.9\columnwidth]{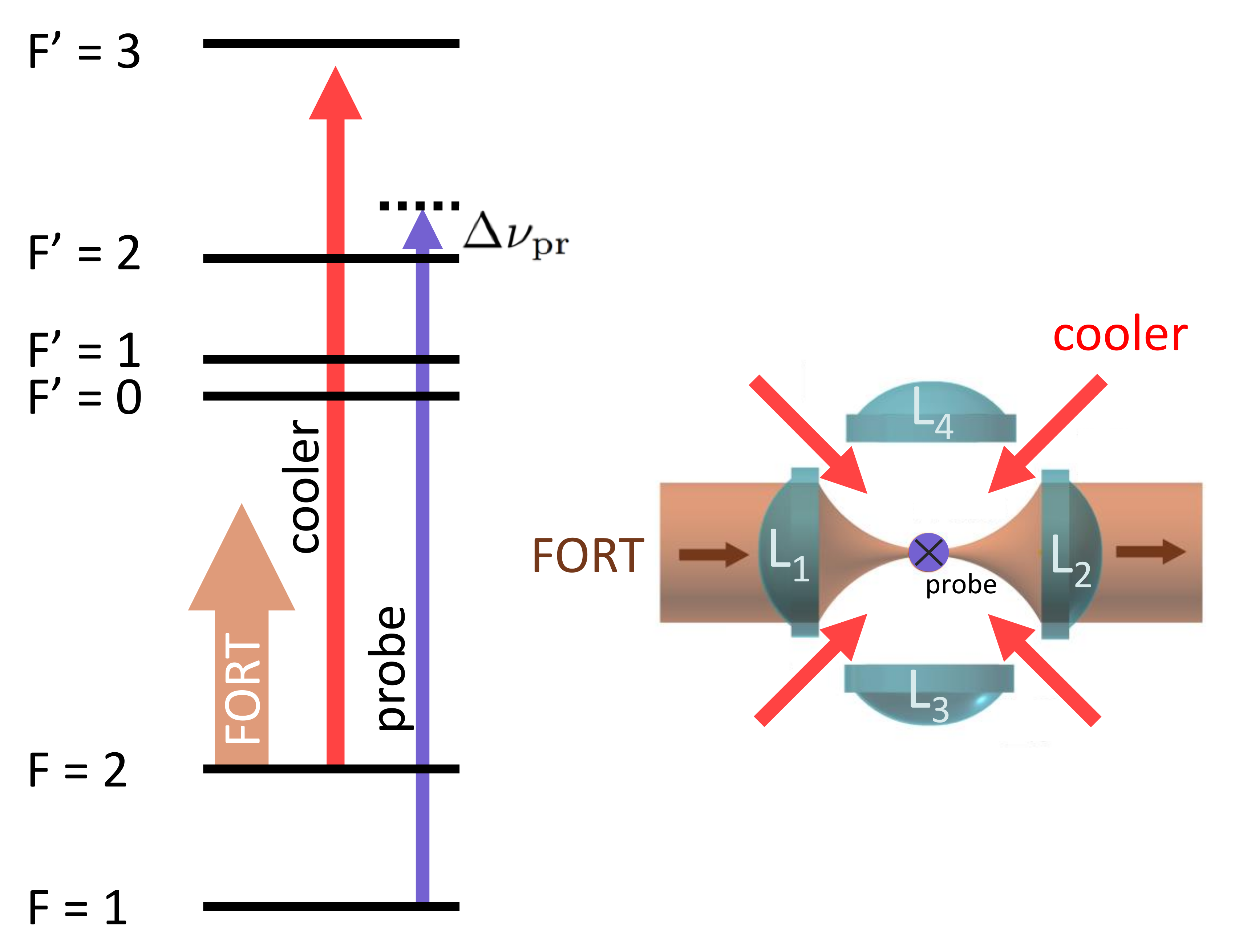} 
\caption{Principle of the quantum jump spectroscopy method.  (Left) Relevant levels of $^{87}$Rb $5S_{1/2} \rightarrow 5P_{3/2}$, D$_2$ transition.  Resonance fluorescence is produced on the closed $2\rightarrow 3'$ hyperfine transition from the $F=2$ ``bright'' state, whereas the $F=1$ ``dark'' state does not fluoresce. {A probe beam with a frequency detuning of $\Delta \nuprobe$} from the $1\rightarrow 2'$ transition can cause Raman transitions to the bright state, resulting in a ``quantum jump'' - a burst of resonance fluorescence that greatly amplifies the effect of the single-photon scattering event that caused the jump. The resonance fluorescence also cools the atom's center of mass motion, returning it to a probe-independent state before the next probe absorption.  
(Right) Geometry of the experiment, viewed from above. Four in-vacuum high numerical-aperture lenses ($L_{1}$ to $L_{4}$) collect resonance fluorescence, and also serve to produce the strongly focused FORT.  Re-pumper propagates together with the four horizontal cooler beams. Circularly-polarized probe beam propagates in the vertical direction {(perpendicular to the plane of the figure)}, {together} with the fifth and sixth cooler beams {(not shown).}
}
\label{fig:principle}
\end{center}
\end{figure}

The method, illustrated in \autoref{fig:principle}, is a spectroscopic probe of the open $1\rightarrow 2'$ transition of the D\textsubscript{2} line, i.e. $5\mathrm{S}_{1/2} F=1\rightarrow 5\mathrm{P}_{3/2} F'=2$  (for brevity, we indicate the ground and excited hyperfine states of this transition with unprimed and primed symbols) that, rather than detecting fluorescence on this transition, detects the induced state change using quantum jump physics, previously studied with ions \cite{SchmidtS2005, MyersonPRL2008}, molecules \cite{BascheN1995}, cavity-bound photons \cite{GleyzesN2007} and quantum dots \cite{VamivakasN2010}. A weak probe beam,  tuned near the $1\rightarrow 2'$ transition, can promote the atom to the $F=2$ ``bright'' ground state by a resonant Raman transition. From there, counter-propagating cooler beams 
drive resonance fluorescence on the closed $2\rightarrow 3'$ transition, Rayleigh scattering hundreds of photons on average before the atom spontaneously falls back to the $F=1$ ``dark'' ground state. The probe and cooler beams are on continuously, so the atom stochastically emits bursts of resonance fluorescence at an average rate set by the rate of $1\rightarrow 2'$ excitation. The probe detuning is scanned across the $1\rightarrow 2'$ line to reveal the ac Stark shifted spectrum of that transition. We refer to this method as \textit{quantum jump spectroscopy}.
 
In this method, resonance fluorescence acts as a high-gain amplifier, scattering many cooler photons for each $1\rightarrow 2'\rightarrow 2$ Raman transition. The amplification gain depends on properties of the cooler light, magnetic fields, trap geometry, and detection efficiencies, all of which can be held constant as the probe frequency is scanned.
In addition, the resonance fluorescence process returns the atom to the $F=1$ state with an internal- and center-of-mass state determined by the resonance fluorescence process, erasing any probe-induced heating or optical pumping. Finally, the $1\rightarrow 2'$ transition has a very small tensor susceptibility.   
\btext{Together, these features reduce systematic effects relative to earlier methods \cite{ShihPRA2013,KaufmanS2014,DeistPRL2022}, leading to an easier data interpretation with more precise results.}

\begin{figure}[t]
\includegraphics[width=0.85\columnwidth]{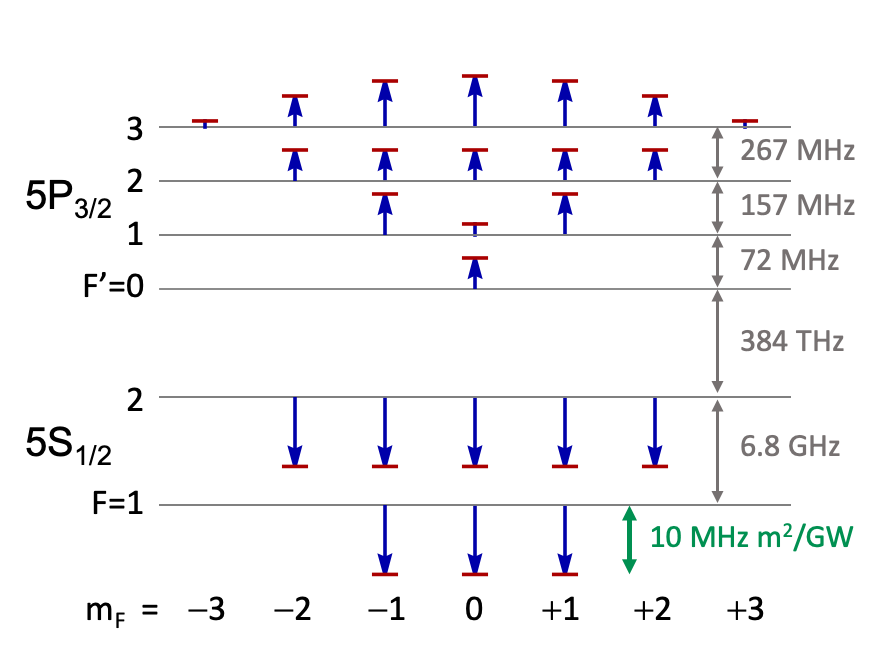}  
\caption{Level shifts for the D\textsubscript{2} line of $^{87}$Rb under linearly-polarized illumination at \SI{852}{\nano\meter},  computed as in \cite{SimonCoopLightshift2017}. Red lines above/below blue arrows show shifted Zeeman sublevels corresponding to the $m_F$ labels below.  Blue arrows show the ac Stark shifts per intensity (scale in green at lower right), relative to the unshifted hyperfine levels (grey horizontal lines). Hyperfine level spacings not to scale. }
\label{fig:LightShifts}
\end{figure}

To demonstrate intensity measurement by quantum jump spectroscopy, we employ a Maltese cross single atom trap \cite{BrunoOE2019, BianchetORE2021}, in which a magneto-optical trap (MOT) with cooler light red-detuned by $4 \gammaNat \approx 2 \pi \times \SI{36}{\mega\hertz}$ from the unshifted $2\rightarrow 3'$ transition is used to load the FORT, and also provides cooler light for the quantum jump spectroscopy. The MOT re-pumper is stabilized near the $1 \rightarrow 1'$ transition.

The FORT light is linearly polarized and stabilized to the Cs D$_2$ line at \SI{852.1}{\nano\meter}, {with an input power of $P_\mathrm{FORT}= \SI{6.8 \pm 0.2}{\milli\watt}$, measured with a power meter before the chamber window. The FORT has an intensity at focus, estimated from the input power and beam waist prior to focusing, of $I_\mathrm{FORT}\supmax \approx \SI{1.6e9}{\watt\per\meter\squared}$ \btext{(and thus trap depth $\approx \SI{740}{\micro \kelvin}$)}\cite{BianchetORE2021}.} As shown in \autoref{fig:LightShifts}, this implies a light shift of $\approx\SI{20}{\mega\hertz}$ on the $1\rightarrow F'$ transitions. \btext{We note that the $F'=2$ state experiences negligible tensor light shifts ($\Delta E^{(2)}_2 / h \approx m_F^2 \times \SI{9.5e-4}{\mega\hertz}$), in comparison with the ones of the $F'= 1$ ($\Delta E^{(2)}_1 / h \approx m_F^2 \times \SI{6.63}{\mega\hertz}$) and $F'= 3$ ($\Delta E^{(2)}_3 / h \approx m_F^2 \times \SI{-2.49}{\mega\hertz}$) states, where $m_F$ is the magnetic quantum number. As a consequence, the transitions to those states are shifted by up to \SI{9.9}{\mega \hertz}, an amount larger that the linewidth of the atomic transition itself. The $1\rightarrow 2'$ transition frequency thus depends on the atom's position, but negligibly on the atom's internal state.}

A circularly-polarized probe beam
with up to
\SI{800}{\nano\watt} of power {in a collimated beam with \SI{2}{\milli\meter} $1/e^2$ diameter} and tunable over \SI{30}{\mega\hertz} on the blue side of the unshifted $1 \rightarrow 2'$ transition with a double-pass AOM, is sent along the downward vertical direction, co-propagating with one of the MOT cooler beams.
Fluorescence is collected by three high-NA lenses ($L_{1}$,$L_{2}$ and $L_{4}$, \btext{henceforth  $L_{i}$}) surrounding the trap center, coupled into single-mode fibers, registered  with \btext{separate} avalanche photodiodes (APDs) and counted in \SI{20}{\milli \second} time bins.

\begin{figure}[t]
\includegraphics[width=1\columnwidth]{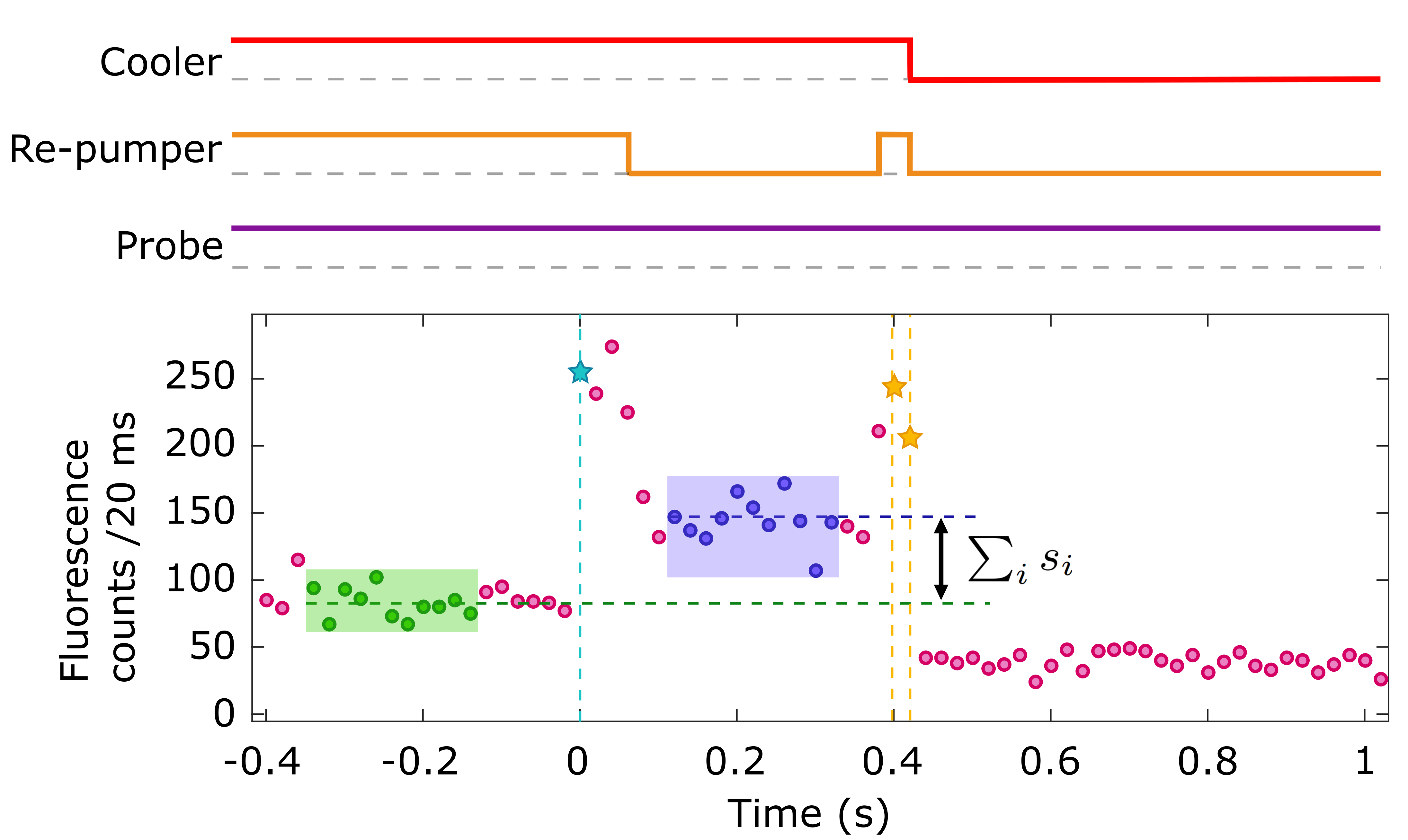}
\caption{Sequence and representative fluorescence signal from a single atom. Upper portion shows time sequence of the cooler, re-pumper and probe light. Dashed lines indicate the respective zero levels. Lower portion shows observed fluorescence counts in \SI{20}{\milli \second} bins, obtained by pooling counts from \btext{Li} collected channels. First detection of an atom in the trap (cyan star) triggers the rest of the sequence, and is taken as the time origin (cyan line). Purple points $d_{i}$ are used to calculate $s_i$ the rate of atom fluorescence collected by channel $i$, green points $b_{i}$ are used to calculate background due to laser scattering and {counts marked with yellow stars ({and} yellow lines) are used to verify atom survival}.}
\label{fig:SingleMeasurement3lenses}
\end{figure}

To acquire fluorescence signals versus probe intensity and versus detuning, we implemented the sequence shown in \autoref{fig:SingleMeasurement3lenses} (upper): starting from an empty FORT, the MOT beams (cooler and re-pumper) are turned on to allow an atom to be trapped. Prior to the atom's arrival, the background count rate is recorded. 

Arrival of an atom is determined when the detected count in channel $L_{1}$ 
is above 
\SI{50}{photons} per bin. 
After this ``trigger'' event, re-pumper and cooler {remain on for  \SI{60}{\milli\second} to cool the trapped atom.}  The re-pumper is then turned off for \SI{300}{\milli\second}, leaving cooler and probe on, to record probe-and-cooler induced fluorescence. The re-pumper is then turned on again for \SI{60}{\milli\second}, to check if the atom is still trapped. The cooler and re-pumper are then turned off. The atom is allowed to leave by turning off the FORT and cooling beams, and the cycle repeats. The probe is on at constant power and frequency during this whole sequence. 

As illustrated in \autoref{fig:SingleMeasurement3lenses} (lower), we record detections $d_{i}$ (shown in purple) and  background counts $b_{i}$ (shown in green) from each trapped atom for a total time of \SI{220}{\milli \second} divided in 11 time bins, where $i$ indicates \btext{collection through lens $L_{i}$}. For any given probe detuning and power, we measure 20 trials like the one shown in \autoref{fig:SingleMeasurement3lenses}. About \SI{65}{\percent} of the atoms survive.  We pool the resulting $d_{i}$ and $b_{i}$ values, to have $N\approx 200$ values of each kind.

\btext{To extract a signal value and error from these data, we assume $b_{i}$ and $d_{i}$ have means $c_i$ and $s_i + c_i$ respectively, where $c_{i}$ is the mean background rate and $s_i$ is the mean atom scattering collected by each channel.  
Thus, $s_i$ is estimated as the sample mean of $\{d_{i} - b_{i}\}$.
We estimate $\sigma_{i,d}^2$ and $\sigma_{i,b}^2$ as the sample variances of $\{d_{i}\}$ and $\{b_{i}\}$, from which we estimate $\sigma_{s_i}^2 = \sigma_{i,d}^2+ \sigma_{i,b}^2$. Error estimates are propagated from these variances.}

\begin{figure}[t]
\includegraphics[width=0.9\columnwidth]{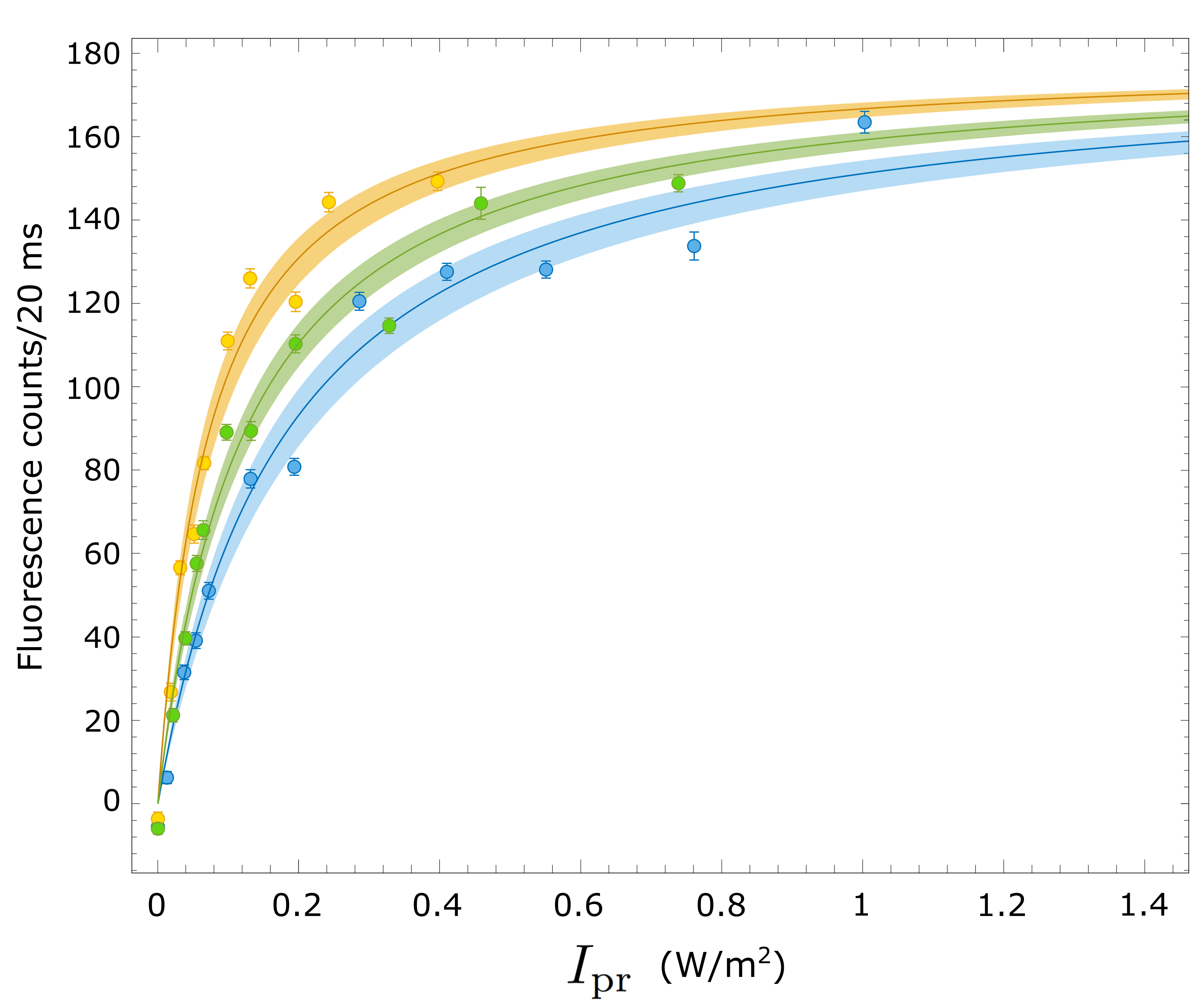}  
\caption{Collected resonance fluorescence rates as a function of probe intensity $I_\mathrm{pr}$ and detuning $\Delta \nuprobe$. Vertical axis shows the net collected signal $\sum_{i} s_i$ \btext{added over $L_{i}$}, where $s_i = \langle d_{i} - b_{i} \rangle$ is averaged over 11 time bins per atom, acquired in \SI{220}{\milli\second} {(\SI{20}{\milli \second} per time bin)}, {and over 11 to 19 atoms.} Upper (orange), middle (green) and lower (blue) curves show detunings $\Delta \nuprobe$ = $\gammaNat/2$, $\gammaNat$ and $3 \gammaNat/2$, respectively, from the frequency  $\nu_{1 \rightarrow 2'}$ of the unshifted $1\rightarrow 2'$ transition.  Error bars show plus/minus one standard error of the mean. Curves show fits with {$\sum_i s_i = \sum_i s\supmax_i \Iprobe/(\Iprobe + I_\mathrm{sat})$}, with
$\sum_{i} s_{i}\supmax=
\SI{179}{}$ counts/\SI{20}{\milli \second} (found by averaging best-fit values for $\sum_{i} s_{i}\supmax$ of the individual detunings)
and best-fit values  {$1/I_\mathrm{sat} = \eta(\nu_{1 \rightarrow 2'} + \Delta \nuprobe)/\gammaJump = \SI{13.6\pm2.2}{\square\meter\per\watt}$, $\SI{8.1\pm1.0}{\square\meter\per\watt}$ and $\SI{5.4\pm0.9}{\square\meter\per\watt}$} for $\Delta \nuprobe = \gammaNat/2$, $\gammaNat$ and $3 \gammaNat/2$, respectively. \btext{Shaded bands show the \SI{95}{\percent} confidence interval.}}
\label{fig:intensitySaturation}
\end{figure}

Representative \btext{pooled} fluorescence \btext{ signals $\sum_{i} s_i$} as a function of probe intensity $\Iprobe$ and detuning $\Delta \nuprobe$ are shown in \autoref{fig:intensitySaturation}. We note that, with $\approx$ 13 atoms per point (requiring less than \SI{6}{\second} of measurement),
the technique resolves detunings in steps of $\gammaNat/2$, where $\gammaNat$ is the natural linewidth of the D\textsubscript{2} transition, and also probe intensity differences of order $\SI{10}{\milli\watt\per\meter\squared}$,  four orders of magnitude below \btext{$I_{0} = \SI{16.69}{\watt\per\meter\squared}$}, the saturation intensity of the D\textsubscript{2} line \cite{Steck2021}.

\newcommand{\PProbe}{P_\mathrm{pr}}
\newcommand{\PSat}{P_\mathrm{sat}}
\newcommand{\IProbe}{I_\mathrm{pr}}
\newcommand{\ISat}{I_\mathrm{sat}}
\newcommand{\Isat}{I_\mathrm{sat}}

\btext{
The fluorescence signals of \autoref{fig:intensitySaturation} show a saturation with intensity that can be understood as follows:
The probe drives the $1\rightarrow 2$ transition with a rate, i.e., probability per unit time, of $R_{1\rightarrow 2} = P_1 \Iprobe \eta(\nuprobe)$, where $P_F$ is the probability to be in state $F$, $\Iprobe$ is the probe intensity, and $\eta(\nuprobe)$ is the efficiency of $1\rightarrow 2$ excitation at probe frequency $\nuprobe$, i.e., the spectral function we seek to measure. The reverse transition happens with rate $R_{2\rightarrow 1} = \Gamma P_2$, where $\Gamma$ depends on the characteristics of the cooler and the $2\rightarrow F'$ transitions, but is independent of the probe.

Defining the saturation power $\Isat \equiv \Gamma/\eta(\nu_\mathrm{pr})$, assuming steady-state, i.e., $R_{1\rightarrow 2} = R_{2\rightarrow 1}$, and a fluorescence emission rate $\propto P_2$, the rate of collected fluorescence via the $i$th channel is 
\begin{equation}
s_i = s_i^{(\mathrm{max})}  \frac{\Iprobe}{\Iprobe+\ISat},
\label{eq:sSaturation}
\end{equation}
where $s_i^{(\mathrm{max})}$ is the atom's maximum fluorescence rate times the channel's collection efficiency.
}

\begin{figure}[t]
\hspace{-2mm}
\includegraphics[width=0.95\columnwidth]{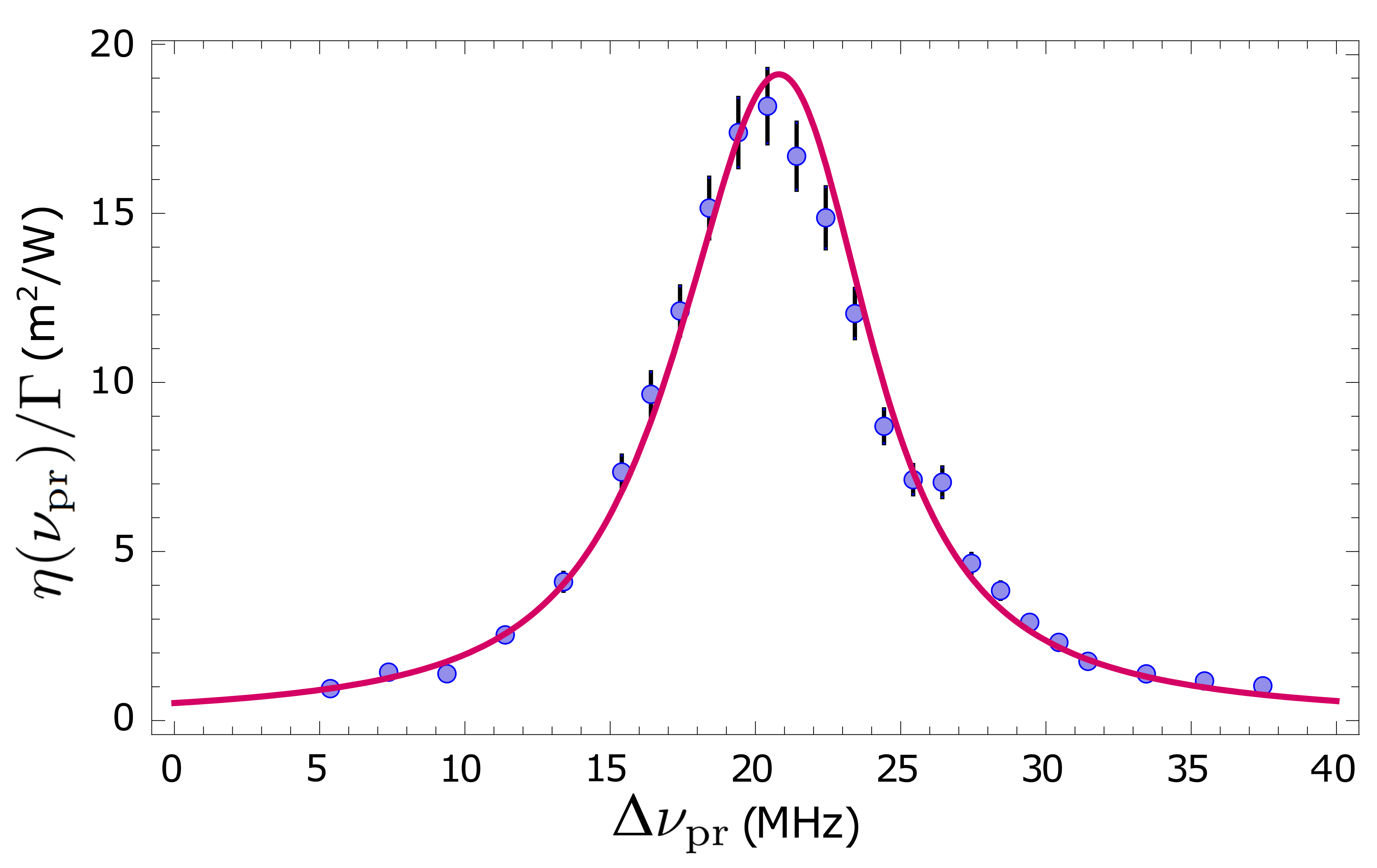}
\caption{Quantum jump spectroscopy of the $5S_{1/2}, F=1 \rightarrow 5P_{3/2}, F'=2$ transition in individual FORT-trapped atoms. Horizontal axis shows detuning $\Delta \nuprobe \equiv \nuprobe-\nu_{1\rightarrow 2'}$ from the unshifted $\nu_{1\rightarrow 2'}$ transition frequency. Vertical axis shows excitation efficiency $\eta(\nuprobe)/\Gamma \btext{= \Isat^{-1}}$ computed via \autoref{eq:sSaturation}. Sequences in which the atom escapes the trap are excluded in post-selection. Each point represents the average of 30 atoms.
Error bars indicate plus/minus one standard {error}.  Background counts, measured with no atom in the trap, have been subtracted. Curves show a fit with \autoref{eq:EtaConvolution}, with FORT intensity at trap center and atom temperature as free parameters.  The fit finds $I_\mathrm{FORT}^{(\mathrm{max})} = \SI{1.593+-0.005e9}{\watt\per\meter\squared}$ at trap center and $T = \SI{36.7+-0.8}{\micro\kelvin}$, with r.m.s. statistical uncertainties found by bootstrapping. These values are in good agreement with \btext{independent estimates by physical optics calculation and release-and-recapture temperature measurement, respectively \cite{BianchetORE2021}.}}
\label{fig:RepumperSpectroscopy}
\end{figure}

\btext{To measure the spectral function $\eta(\nuprobe)$, we first measure fluorescence $s_i$ versus $\Iprobe$ and fit with \autoref{eq:sSaturation} to obtain values for $s_i\supmax$, as shown in  \autoref{fig:intensitySaturation}.
We then set a probe frequency $\nu_\mathrm{pr}$, adjust $\Iprobe$ to achieve $s_i/s_i\supmax \approx 1/3$ (a condition that minimizes statistical uncertainty in the spectral function),  record  $\Iprobe$ and $s_i$ for 30 trials, average weighted by $\sigma_{s_i}^{-2}$, and compute $\eta(\nu_\mathrm{pr}) = \Isat^{-1}$ using \autoref{eq:sSaturation}. Repeating for a range of $\nu_\mathrm{pr}$ we obtain spectra such as that shown in \autoref{fig:RepumperSpectroscopy}. We note that line broadening due to saturation is automatically compensated in this method, and that the probe intensity is always 
well below the $1\rightarrow 2'$ saturation intensity.
}

Via light shifts, this spectrum gives information on both the FORT intensity and thus beam shape, and on the atomic center-of-mass spatial distribution and thus atom temperature. If the probe laser instantaneous frequency is $\nu$, and the instantaneous light-shift is $\Delta_{eg} = (\delta_{e} - \delta_g)I_\mathrm{FORT}(\mathbf{x})$, where $\delta_e$, $\delta_g$ are the per-intensity light shifts of the excited and ground state, respectively, and $\mathbf{x}$ is the instantaneous position of the atom, then the instantaneous {efficiency} of excitation is  
\begin{eqnarray}
\eta & \propto & f_\mathrm{nat}(\nu_{eg} + \Delta_{eg}-\nu),
\end{eqnarray}
where $\nu_{eg} = (E_e-E_g)/2\pi\hbar$ is the unshifted line center 
and $f_\mathrm{nat}(\nu) \propto  1/[(\gammaNat/2)^2 + (2 \pi \nu)^2]$ is the natural line shape function. Averaging over the distribution of light shifts $f_\Delta(\Delta_{eg})$, and the probe laser's line-shape function $f_\mathrm{pr}(\delta)$, with $\delta \equiv \nu-\nuprobe$, we obtain  
\begin{eqnarray}
\label{eq:EtaConvolution}
\eta(\nuprobe) & \propto & \int d\delta\,d\Delta_{eg}\, f_\mathrm{nat}(\nu_{eg} + \Delta_{eg}-\nuprobe -\delta) \nonumber \\ & & \times f_\mathrm{pr}(\delta) f_\Delta(\Delta_{eg}),
\end{eqnarray}
i.e., the convolution of $f_\mathrm{nat}(\nu)$ with $f_\mathrm{pr}(\delta)$ and $f_\Delta(-\Delta_{eg})$. 

To relate this to the atom temperature, we note that the optical potential is  $V = \alpha \Delta_{eg}$, where $\alpha \equiv 2\pi\hbar \delta_g/(\delta_e-\delta_g)$.  Assuming the atom's center-of-mass coordinate is thermally distributed, $f_\Delta$ is given by a Boltzmann distribution $f_\Delta \propto \exp[-\beta V] \rho(V)$, where $\beta \equiv 1/k_B T$ and $\rho(V)$ is the potential density of states.  If the potential is quadratic with minimum $V_\mathrm{min}$, this gives
\begin{eqnarray}
\label{eq:FDelta}
f_\Delta(\Delta) &\propto&  \sqrt{\alpha \Delta - V_\mathrm{min}} e^{- \beta(\alpha \Delta - V_\mathrm{min})} \beta^{3/2}
\end{eqnarray} 
for $\alpha \Delta > V_\mathrm{min}$, and zero otherwise \footnote{\btext{We note that $f_\Delta \propto \exp[-\beta V]\rho(V)$ depends on $V_\mathrm{min}$ and on $T$, but not on derivatives $\partial_i V$ or $\partial_i \partial_j V$, $i, j \in \{x,y,z\}$. This makes the result insensitive to details of the potential shape, provided the temperature is well below the trap depth. We have also performed the analysis assuming $I_\mathrm{FORT}(\mathbf{x})$ describes an ideal gaussian beam, and found $I_\mathrm{FORT}^{(\mathrm{max})}$ and $T$ that agree with the quadratic approximation to within  statistical uncertainty.}}. 

The line center reflects the average light shift, which depends strongly on  the maximum intensity $I_\mathrm{FORT}^{(\mathrm{max})} \equiv \max_{\mathbf{x}} I_\mathrm{FORT}(\mathbf{x})$ and weakly on the temperature $T$, whereas the line width depends more strongly on $T$. Fitting the data of \autoref{fig:RepumperSpectroscopy}, and using a bootstrapping procedure to estimate the fitting uncertainties, we find $I_\mathrm{FORT}^{(\mathrm{max})} = \SI{1.593+-0.005e9}{\watt\per\meter\squared}$ and a temperature $T = \SI{36.7+-0.8}{\micro\kelvin}$. The reduced $\chi^2$ of this fit is 3.8, which suggests that there are other perturbations roughly comparable to these statistical uncertainties.
Relating the obtained value of $I_\mathrm{FORT}^{(\mathrm{max})}$ with the waist $w$ of the FORT beam, defined as the $1/e^2$ radius of intensity, and assuming the Gaussian beam relation $I_\mathrm{FORT}^{(\mathrm{max})} = 2 P_\mathrm{FORT}/\pi w^2$ with FORT power $P_\mathrm{FORT}= {\SI{6.8 \pm 0.2}{\milli\watt}}$, the waist is $w$=\SI{1.65\pm0.02}{\micro\meter}, in good agreement with prior estimates \cite{BianchetORE2021}. The implied r.m.s.~width of the center-of-mass distribution is \SI{0.184}{\micro\meter} in the radial directions and  \SI{1.58}{\micro\meter} in the longitudinal, so the atom samples the FORT intensity distribution with sub-wavelength transverse resolution.

\btext{Although a full systematic error analysis is beyond the scope of this work, we note that the Zeeman shift of the $1\rightarrow 2'$ transition is $B (\gamma_{2'}m' - \gamma_{1}m)/2\pi$, where $B$ is the magnetic field strength, $\gamma_1/2\pi =  \SI{-0.7}{\mega\hertz\per\gauss}$ and  $\gamma_{2'}/2\pi = \SI{0.93}{\mega\hertz\per\gauss}$ are the $F=1$ and $F'=2$ gyromagnetic ratios, respectively, and $m$, $m'$ are the corresponding magnetic quantum numbers \cite{Steck2021}. The magnitude of the transition shift is thus at most $B \times \SI{2.56}{\mega\hertz\per\gauss}$.  The measured magnetic field fluctuations of the laboratory are $B \lesssim  \SI{10}{\milli\gauss}$ \cite{BehboodAPL2013}, implying line shifts and broadening due to Zeeman shifts below \SI{26}{\kilo\hertz}.  Using a pulsed probe synchronized to the ac power line could reduce this by more than an order of magnitude \cite{BehboodAPL2013}. Vector light shifts due to elliptically-polarized FORT light can be analyzed similarly. The shift is $\Delta I(\gamma^{(\mathrm{opt})}_{2'} m' - \gamma^{(\mathrm{opt})}_1 m)$, where $\Delta I$ is the difference in intensity between $\sigma_+$ and $\sigma_-$ polarization components (quantization axis along the FORT propagation direction), and
$\gamma^{(\mathrm{opt})}_1 = \SI{-4.1e-10}{\mega\hertz\meter\squared\per\watt} $ and $\gamma^{(\mathrm{opt})}_{2'} = \SI{-1.49e-9}{\mega\hertz\meter\squared\per\watt}$ are the calculated vector light shift coefficients  for our FORT wavelength \cite{SimonCoopLightshift2017}. In our geometry, the circularly-polarized probe drives simultaneously the $m'=m$  and $m'=m \pm 1$ transitions in the ratio 2:1:1, implying a transition-averaged shift of  at most $\Delta I \times \SI{1.08e-9}{\mega\hertz\meter\squared\per\watt}$. Assuming $\Delta I = 0.013 \times I_\mathrm{FORT}^{(\mathrm{max})} = \SI{2.0e7}{\watt\per\square\meter}$, corresponding to the maximum $\Delta I$ of a beam with linear polarization extinction ratio $10^5$:1, we find a maximum vector light shift of \SI{22}{\kilo\hertz} for the transition. 
For comparison, the \SI{5e6}{\watt\per\square\meter} statistical uncertainty of $I_\mathrm{FORT}^{(\mathrm{max})}$ corresponds to a scalar transition light shift of \SI{70}{\kilo\hertz}. Line broadening is quadratic in the dispersion of such shifts, and is negligible here. 
}

We note possible extensions of the technique: \btext{first, the method could be implemented step-wise, with sequential state preparation, probing, and readout. This would remove noise associated with the stochastic $1\leftrightarrow 2$ jumps in the continuous implementation. Second, circularly- or elliptically-polarized fields could be measured without state-dependent shifts, if the atom is optically-pumped to a specific $F=1, m_F$ state \cite{TeyNP2008}. Third,}
the spectroscopy could be performed with the probe tuned to the $1\rightarrow 1'$ transition. On this transition, the excitation efficiency will be strongly Zeeman-state and probe-polarization dependent, due to selection rules and strong tensor light shifts of the $F'=1$ state, as shown in \autoref{fig:LightShifts}. These features enable internal-state-selective detection with the same advantages of high gain and low perturbation that we have demonstrated using the $1\rightarrow 2'$ transition. 

\noindent\textit{Conclusion --} We have proposed and demonstrated the use of a single neutral \textsuperscript{87}Rb atom for precision, sub-wavelength sensing of optical intensity, implemented by a quantum jump spectroscopy technique.  A very low intensity probe near the $F=1 \rightarrow F'=2$ hyperfine transition of the D\textsubscript{2} line drives ``quantum jumps,'' i.e., resonant Raman transitions, into the $F=2$ ground state. A second laser near the $F=2 \rightarrow F'=3$ cycling transition induces a burst of resonance fluorescence for each Raman transition, greatly amplifying the detectable signal. By scanning the probe frequency, the spectrum of $F=1 \rightarrow F'=2$ excitation is measured, indicating the distribution of ac Stark shifts on this transition, which suffers negligible broadening from tensor light shifts. From this spectrum we obtain the intensity at trap centre and the atom's temperature.
The technique can be extended to perform Zeeman-state-selective readout.

We thank Enes Aybar for helpful discussions. This work was supported by H2020 Future and Emerging Technologies Quantum Technologies Flagship projects macQsimal (Grant Agreement No. 820393) and  QRANGE (Grant Agreement No.  820405);   Marie Sk{\l}odowska-Curie grant agreement No. 847517,
Spanish Ministry of Science projects OCARINA (Grant No. PGC2018-097056-B-I00), 17FUN03 USOQS, which has received funding from the EMPIR programme co-financed by the Participating States and from the European Union's Horizon 2020 research and innovation programme; La Caixa Foundation under agreement [LCF/BQ/SO15/52260044]; and ``Severo Ochoa'' Center of Excellence CEX2019-000910-S
Generalitat de Catalunya through the CERCA program; Ag\`{e}ncia de Gesti\'{o} d'Ajuts Universitaris i de Recerca Grant No. 2017-SGR-1354;
{Grant PRE2020-094392 financed by MCIN/AEI/10.13039/501100011033 and FSE “El FSE invierte en tu futuro”};
Secretaria d'Universitats i Recerca del Departament d'Empresa i Coneixement de la Generalitat de Catalunya, co-funded by the European Union Regional Development Fund within the ERDF Operational Program of Catalunya (project QuantumCat, ref. 001-P-001644); Fundaci\'{o} Privada Cellex; Fundaci\'{o} Mir-Puig. 

\bibliographystyle{apsrev4-1no-url}
\bibliography{QJS_PRR}


\end{document}